\documentclass[letterpaper]{article}
\usepackage{ijcai17}
\usepackage{times}
\usepackage{helvet}
\usepackage{courier}
\usepackage{color}

\usepackage{verbatim}
\usepackage{amsmath}
\usepackage{amssymb}
\usepackage{amsthm}
\usepackage{rotating}
\usepackage{algorithm}
\usepackage[noend]{algpseudocode}
\usepackage{siunitx}

\usepackage{tabularx}
\usepackage{subfigure}
\usepackage{graphicx}
\usepackage{url}
\usepackage{multirow}
\usepackage[font=small,skip=1pt]{caption}

\usepackage{silence}
\newtheorem{fundamental}{Fundamental Rule of Poker Strategy}

\begin{document}
\title{Computing Human-Understandable Strategies}
\author{Sam Ganzfried and Farzana Yusuf\\
School of Computing and Information Sciences\\
Florida International University\\
\{sganzfri@cis.fiu.edu, fyusu003@fiu.edu\}
}

\maketitle
\begin{abstract}
Algorithms for equilibrium computation generally make no attempt to ensure that the computed strategies are understandable by humans. For instance the strategies for the strongest poker agents are represented as massive binary files. In many situations, we would like to compute strategies that can actually be implemented by humans, who may have computational limitations and may only be able to remember a small number of features or components of the strategies that have been computed. We study poker games where private information distributions can be arbitrary. We create a large training set of game instances and solutions, by randomly selecting the information probabilities, and present algorithms that learn from the training instances in order to perform well in games with unseen information distributions. We are able to conclude several new fundamental rules about poker strategy that can be easily implemented by humans.
\end{abstract}

\section{Introduction}
\label{se:intro}
Large-scale computation of strong game-theoretic strategies is important in many domains. For example, there has been significant recent study on solving game-theoretic problems in national security from which real deployed systems have been built, such as a randomized security check system for airports~\cite{Paruchuri08:Playing}. Typically large-scale equilibrium-finding algorithms output massive strategy files (which are often encoded in binary), which are stored in a table and looked up by a computer during gameplay. For example, creators of the optimal strategy for two-player limit Texas hold 'em recently wrote, ``Overall, we require less than 11 TB of storage to store the regrets and 6 TB to store the average strategy during the computation, which is distributed across a cluster of computation nodes. This amount is infeasible to store in main memory,...''~\cite{Bowling15:Heads-up}. While such approaches lead to very strong computer agents, it is difficult to see how a human could implement these strategies. For cases where humans will be making real-time decisions we would like to compute strategies that are easily interpretable.  

Suppose a human plans to play the following two-player no-limit poker game. Player 1 and player 2 both ante \$0.50 and are dealt a card from a 10-card deck and each have a stack of \$3 after posting the ante. Player 1 can bet any multiple of 0.1 from 0 to 3 (he has 31 possible actions for each hand). Player 2 can then call or fold. If player 2 folds, then player 1 wins the \$1 from the antes. Otherwise the player with the better card wins the amount bet plus the antes. For example, if player 1 has a 4, player 2 has a 9, player 1 bets 0.4 and player 2 calls, then player 2 wins 0.4 plus the antes.

\begin{figure}[!t]
\centering
\begin{minipage}{.43\textwidth}
\centering
\includegraphics[width=0.82\linewidth]{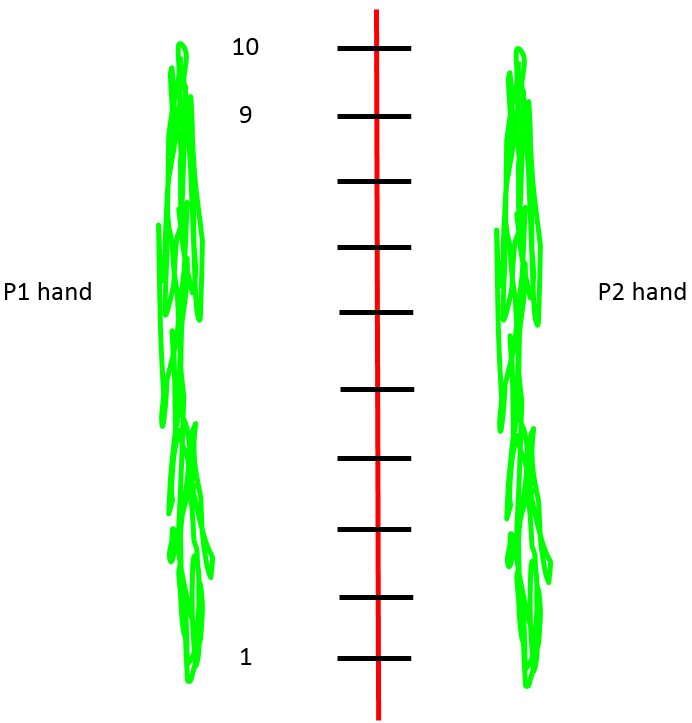}
\caption{Both players are dealt private information uniformly at random over all hands.}
\label{fi:uniform}
\end{minipage}
\hfill
\centering
\begin{minipage}{.43\textwidth}
\centering
\includegraphics[width=0.82\linewidth]{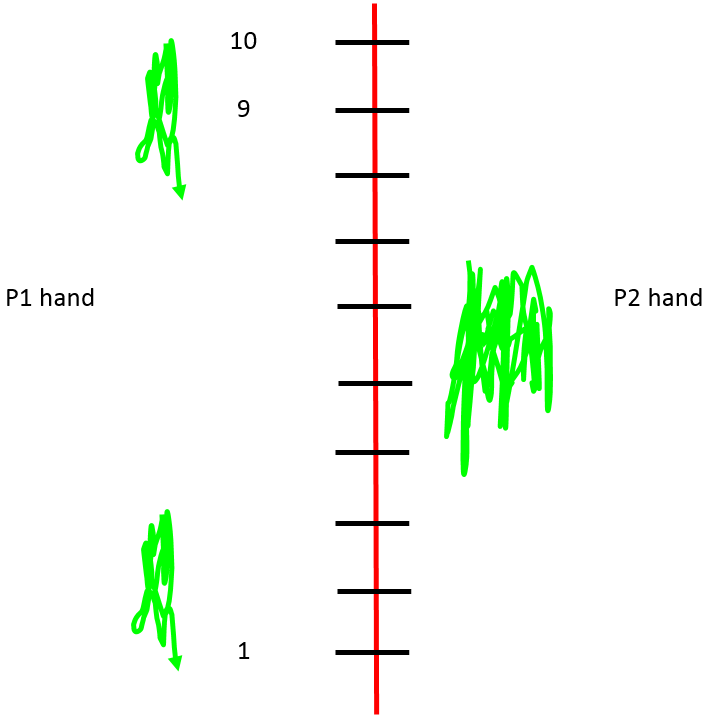}
\caption{Player 1 is dealt very strong or weak hand and player 2 is always dealt mediocre hand.}
\label{fi:p1polar}
\end{minipage}
\hfill
\centering
\begin{minipage}{.43\textwidth}
\centering
\includegraphics[width=0.82\linewidth]{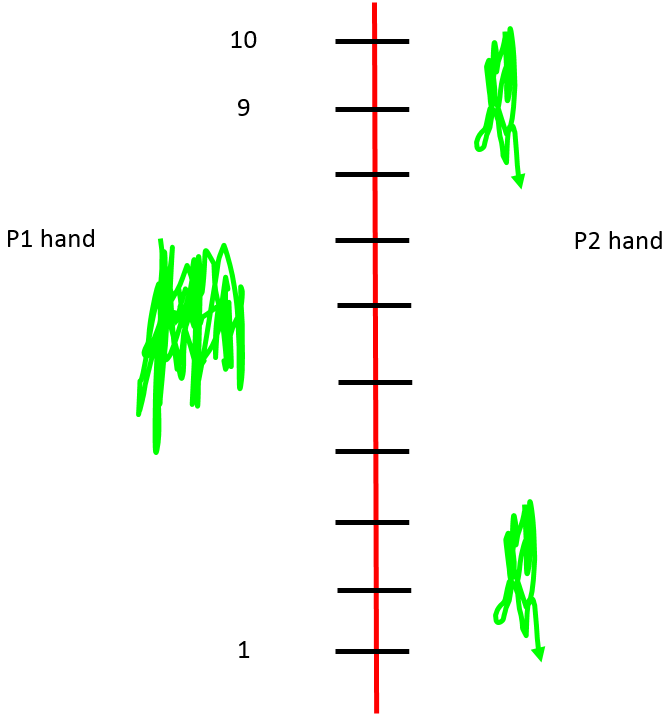}
\caption{Player 1 is always dealt mediocre hand and player 2 is dealt very strong or weak hand.}
\label{fi:p2polar}
\end{minipage}
\end{figure}

If both players are dealt cards uniformly at random (Figure~\ref{fi:uniform}), then a Nash equilibrium strategy for player 1 is:
\begin{itemize}
\item Card 1: Bet 0.1 pr. 0.091, 0.6 pr. 0.266, 1.8 pr. 0.643 
\item Card 2: Bet 0 pr. 0.660, 0.3 pr. 0.231, 0.6 pr. 0.109 
\item Card 3-6: Bet 0 pr. 1 
\item Card 7: Bet 0.1 pr. 1 
\item Card 8: Bet 0.3 pr. 1 
\item Card 9: Bet 0.6 pr. 1 
\item Card 10: Bet 1.8 pr. 1
\end{itemize}
This can be computed quickly using, e.g., a linear programming formulation~\cite{Koller92:Complexity}. 

However, suppose the cards are dealt according a different distribution: player 1 is either dealt a very strong hand (10) or a very weak hand (1) with probability 0.5 while player 2 is always dealt a medium-strength hand (Figure~\ref{fi:p1polar}). Then the equilibrium strategy for player 1 is:
\begin{itemize}
\item Card 1: Bet 0 pr. 0.25, 3 pr. 0.75 
\item Card 10: Bet 3 pr. 1
\end{itemize}
If player 1 is always dealt a medium-strength hand (5) while player 2 is dealt a very strong or very weak hand with probability 0.5 (Figure~\ref{fi:p2polar}), then the equilibrium strategy is:
\begin{itemize}
\item Card 5: Bet 0 pr. 1 
\end{itemize}

What if player 1 is dealt a 1 with probability 0.09, 2 with probability 0.19, 3 with probability 0.14, etc.? For each game instance induced by a probability distribution over the private information, we could solve it quickly if we had access to an LP solver. But what if a human is to play the game without knowing the distribution in advance and without aid of a computer? He would need to construct a strong game plan in advance that is capable of playing well for a variety of distributions with minimal real-time computation. A natural approach would be to solve and memorize solutions for several games in advance, then quickly determine which of these games is closest to the one actually encountered in real time. This is akin to the $k$-nearest neighbors ($k$-nn) algorithm from machine learning. A second would be to construct understandable rules (e.g., if .. else ..) from a database of solutions that can be applied to a new game. This is akin to the decision tree and decision list approaches. Thus, we are proposing to apply approaches from machine learning in order to improve human ability to implement Nash equilibrium strategies. Typically algorithms from machine learning have been applied to game-theoretic agents only in the context of learning to exploit mistakes of suboptimal opponents (aka opponent exploitation). By and large the approaches for computing Nash equilibrium and opponent exploitation have been radically different. We provide a new perspective here by integrating learning into the equilibrium-finding paradigm.

We present a novel learning formulation of this problem. In order to apply algorithms we develop novel distance functions (both between pairs of input points and between pairs of output points) which are more natural for our setting than standard distance metrics. To evaluate our approaches we compute a large database of game solutions for random private information distributions.  We are able to efficiently apply $k$-nn to the dataset using our custom distance functions. We observed that we are able to obtain low testing error even when training on a relatively small fraction of the data, which suggests that it is possible for humans to learn strong strategies by memorizing solutions to a carefully selected small set of presolved games. However, this approach would require humans to quickly be able to compute the distance between a new game and all games from the training database in order to determine the closest neighbor, which could be computationally taxing. Furthermore, there are some concerns as to whether this would actually constitute ``understanding'' as opposed to ``memorizing.'' Thus, we focus on the decision tree approach, which allows us to deduce simple human-understandable rules that can be easily implemented. 


While prior approaches for learning in games of imperfect information (and poker specifically) typically utilize many domain-specific features (e.g., number of possible draws to a flush, number of high cards on the public board, etc.), we prefer to develop approaches that are more robust and do not require knowing expert domain features (since they are likely not relevant for other domains and, in the case of poker, may not be relevant even for other seemingly similar variants). The features we use are the cumulative distribution function values of the private information states of the players, which are based purely on the rules of the game. (We also compare performance of using several other representations, e.g., using pdf values, and separating out the data for each hand to create 10 data points per game instance instead of 1.) Thus, the approach is general and not reliant on expert poker knowledge.

The problem of constructing human-interpretable rules has been studied recently in machine learning, e.g.,~\cite{Bertsimas11:Ordered,Jung17:Simple}, particularly for medical applications~\cite{Lakkaraju16:Interpretable,Letham15:Interpretable,Marewski12:Heuristic}.

\section{Qualitative models and endgame solving}
\label{se:qual}
\begin{figure}[!t]
\centering
\begin{minipage}{.47\textwidth}
\centering
\includegraphics[width=0.95\linewidth]{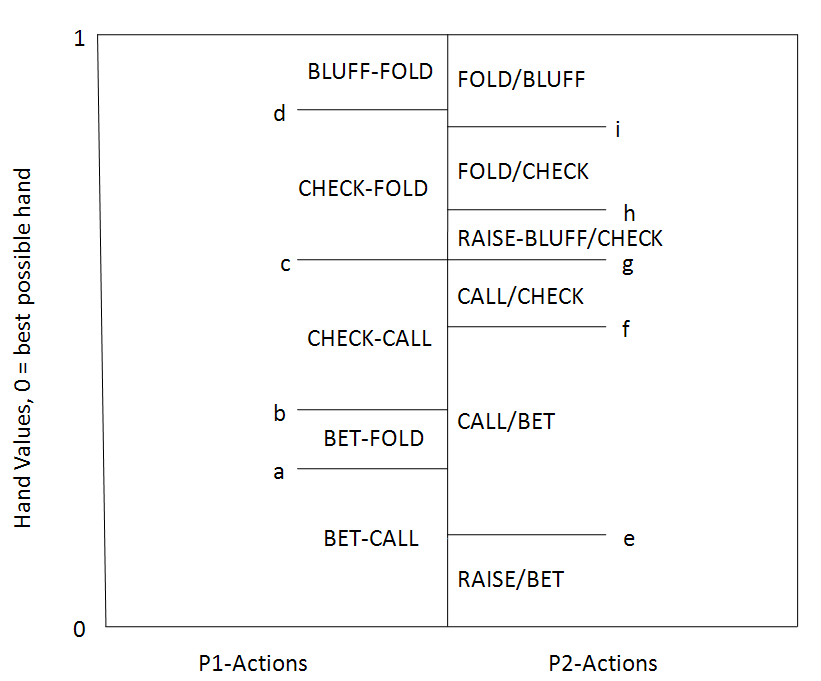}
\caption{First qualitative model for two-player limit Texas hold 'em river endgame play.}
\label{fi:param1}
\end{minipage}
\hfill
\centering
\begin{minipage}{.47\textwidth}
\centering
\includegraphics[width=0.95\linewidth]{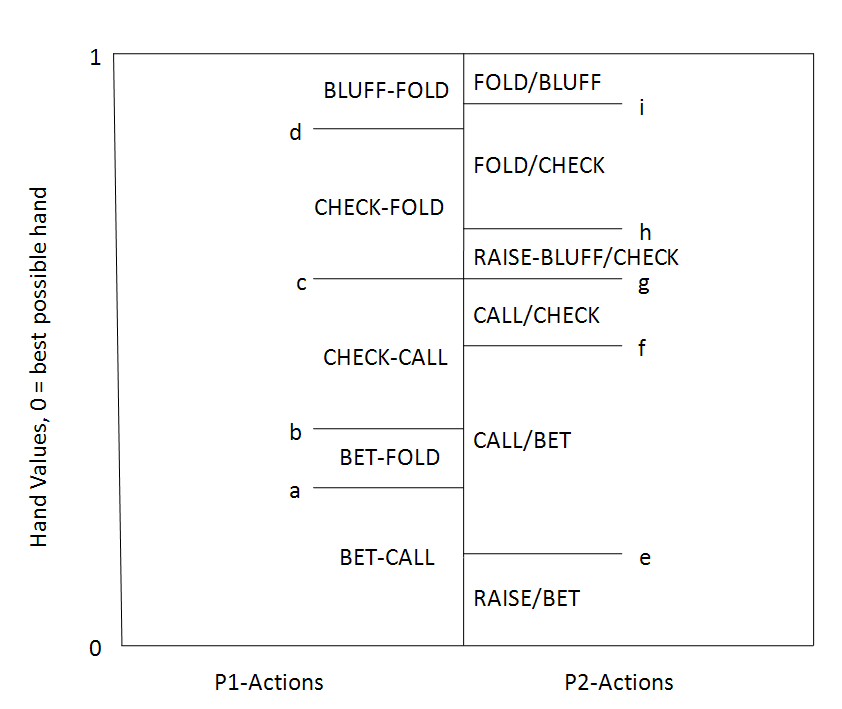}
\caption{Second qualitative model for two-player limit Texas hold 'em river endgame.}
\label{fi:param2}
\end{minipage}
\hfill
\centering
\begin{minipage}{.47\textwidth}
\centering
\includegraphics[width=0.95\linewidth]{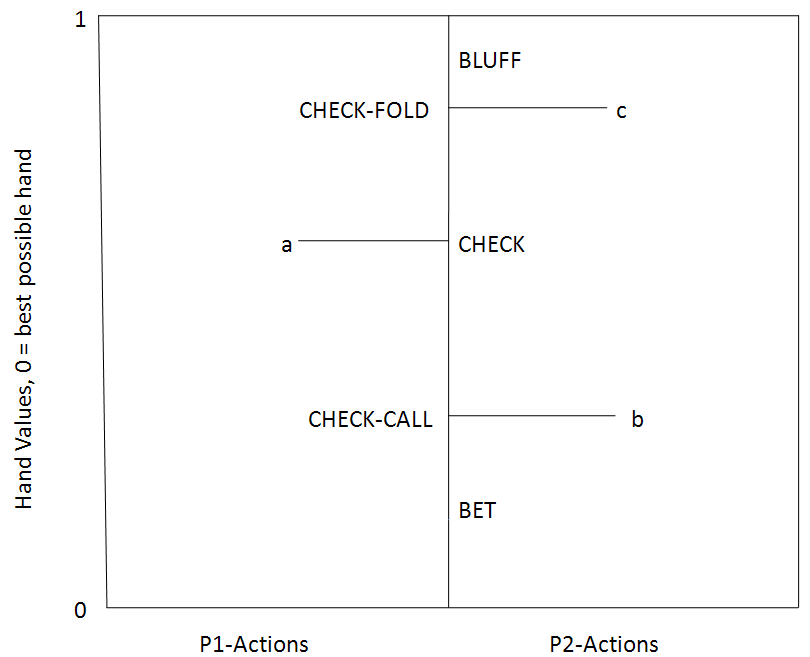}
\caption{Third qualitative model for two-player limit Texas hold 'em river endgame.}
\label{fi:param3}
\end{minipage}
\end{figure}

There has been some prior study of human understandable strategies in imperfect-information games, and in poker specifically. Ankenman and Chen compute analytical solutions of several simplified poker variants by first assuming a given human-understandable qualitative structure on the equilibrium strategies, and then computing equilibrium strategies given this presumed structure, typically by solving a series of indifference equations~\cite{Ankenman06:Mathematics}. While the computed strategies are generally interpretable by humans, the models were typically constructed from a combination of trial and error and expert intuition, and not constructed algorithmically. More recent work has shown that leveraging such qualitative models can lead to new equilibrium-finding algorithms that outperform existing approaches~\cite{Ganzfried10:Computing}. That work proposed three different qualitative models for the final round endgame of two-player limit Texas hold 'em (Figures~\ref{fi:param1}--~\ref{fi:param3}), and showed empirically that endgame equilibrium strategies conformed to one of the models for all input information distributions (and that all three were needed). Again here the models were constructed by manual trial and error, not learned algorithmically.

We note that while the problem we are considering in this paper is a ``toy game,'' it captures important aspects of real poker games and we expect our approaches to have application to larger more realistic variants. In the recent Brains vs. Artificial Intelligence two-player no-limit Texas hold 'em competition, the agent Claudico computed the strategy for the final betting round in real time, and the best human player in the world for that variant (Doug Polk) commented that the ``endgame solver'' was the strongest component of the agent~\cite{Ganzfried15:Reflections}, and endgame solving was also a crucial component of subsequent success of the improved agent Libratus~\cite{Brown17:Safe}. The creator of another recent superhuman agent has stated that ``DeepStack is all endgame solving,'' referring to the fact that its algorithm works by viewing different rounds of the game as separate endgames which are solved independently, using deep learning to estimate the values of the endgames terminal states~\cite{Moravcik17:DeepStack}. Endgame solving assumes that both agents had private information distributions induced by the strategies for the prior rounds using Bayes' rule, assuming they had been following the agent's strategy for the prior rounds~\cite{Ganzfried15:Endgame}. The game we study here is very similar to no-limit Texas hold 'em endgames, except that we are assuming a ten-card deck, specific stack sizes and betting increment, and that raises are not allowed. We expect our analysis to extend in all of these dimensions and that our approaches will have implications for no-limit Texas hold 'em strategy. No-limit Texas hold 'em is the most popular poker variant for humans, and is a widely recognized AI challenge problem. The game tree has approximately $10^{165}$ states for the variant played in the AAAI Annual Computer Poker Competition~\cite{Johanson13:Measuring}. There has been significant interest in endgame solving in particular in the last several years, and several new advances have been developed~\cite{Burch14:Solving,Moravcik16:Refining,Brown17:Safe,Moravcik17:DeepStack}. 

\section{Learning formulation}
\label{se:learning}
We now describe how we formulate the problem of computing a solution to a new game instance from a database of solutions to previously solved game instances as a learning problem. The inputs to the learning problem will be the 20 values of the private information cumulative distribution function (cdf). First are the ten values for player 1 (the probability he is dealt $\leq$ 1, probability he is dealt $\leq$ 2, etc.), followed by the ten cdf values for player 2. For example for the uniform case the input would be 
$$X = (0.1, 0.2, 0.3, 0.4, 0.5, 0.6, 0.7, 0.8, 0.9, 1,$$
$$0.1, 0.2, 0.3, 0.4, 0.5, 0.6, 0.7, 0.8, 0.9, 1),$$
for the situation where player 1 is dealt a 10 or 1 with probability 0.5 and player 2 is always dealt a 5 it is
$$X = (0.5, 0.5, 0.5, 0.5, 0.5, 0.5, 0.5, 0.5, 0.5, 1,$$
$$0, 0, 0, 0, 1, 1, 1, 1, 1, 1),$$
and for the situation where player 1 is always dealt a 5 and player 2 is dealt a 10 or 1 with probability 0.5 it is
$$X = (0, 0, 0, 0, 1, 1, 1, 1, 1, 1,$$ 
$$0.5, 0.5, 0.5, 0.5, 0.5, 0.5, 0.5, 0.5, 0.5, 1).$$

The output will be a vector of the 310 Nash equilibrium strategy probabilities of betting each size with each hand. First for betting 0, 0.1, 0.2, \ldots, 3 with 1, then with 2, etc. (recall that there are 31 sizes for each of ten hands). For example for the uniform case the output would be
$$y = (0, 0.091, 0, 0, 0, 0, 0.266, 0, 0, 0, 0, 0, 0, 0, 0,0, 0,$$ 
$$0, 0.643, 0, 0, 0, 0, 0, 0, 0, 0, 0, 0, 0, 0,\ldots).$$
We could have created ten different data points for each game corresponding to the strategy for each hand, as opposed to predicting the full strategy for all hands; however we expect that predicting complete strategies is better than just predicting strategies for individual hands because the individual predicted hand strategies may not balance appropriately and could be highly exploitable as a result. We will explore this design choice in the experiments in Section~\ref{se:experiments}.

\begin{figure}[!ht]
\centering
\includegraphics[scale=0.41]{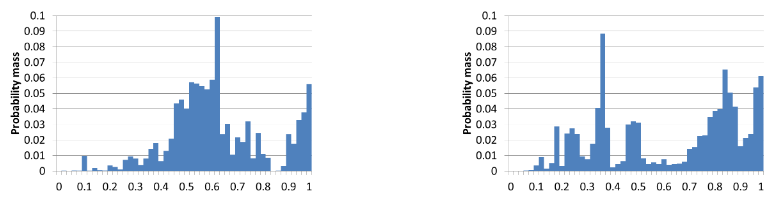}
\caption{Earth-mover's distance has proven a successful metric for distance between probability distributions.}
\label{fi:histograms}
\end{figure}

To perform learning on this formulation, we need to select a distance function to use between a pair of inputs as well as a distance (i.e., cost) between each pair of outputs. Standard metrics of Euclidean or Manhattan distance are not very appropriate for probability distributions. A more natural and successful distance metric for this setting is earth mover's distance (EMD). While early approaches for computing groupings of hands used L2~\cite{Gilpin07:Potential}, EMD has been shown to significantly outperform other approaches, and the strongest current approaches for game abstraction use EMD~\cite{Johanson13:Evaluating}. Informally, EMD is the ``minimum cost of turning one pile into the other, where the cost is assumed to be amount of dirt moved times the distance by which it is moved,'' and there exists a linear-time algorithm for computing it for one-dimensional histograms (Figure~\ref{fi:histograms}).

We define a new distance metric for our setting that generalizes EMD to multiple distributions. Suppose we want to compute the distance between training input vector $X$ and testing input vector $\hat{X}$. Each vector contains 20 probabilities, 10 corresponding to player 1's distribution and 10 to player 2's. Our distance function will compute the EMD separately for each player, then return the average (Algorithm~\ref{al:distance-input}). We note that before the aggregation we normalize the EMD values by the maximum possible value (the distance between a point mass on the left-most and right-most columns) to ensure that the maximum of each is 1. We also create a new distance (i.e., cost) function between predicted output vector $\hat{Y}$ and the actual output vector from the training data $Y$ (the output vectors have length 310, corresponding to 31 bet sizes for 10 hands). It computes EMD separately for the strategy vectors of size 31 for each hand which are then normalized and averaged (Algorithm~\ref{al:distance-output}). After specifying the form of the inputs and outputs and a distance metric between each pair of inputs and outputs, we have formulated the problem as a machine learning problem.

\begin{algorithm}[!ht]
\small
\caption{Distance between input vectors $X$, $\hat{X}$}
\label{al:distance-input} 
\textbf{Inputs}: cdf vectors $X, \hat{X}$, number of players $n$, deck size $d$
\begin{algorithmic}
\State $X'$ $\gets$ cdf-to-pdf($X$); $\hat{X'}$ $\gets$ cdf-to-pdf($\hat{X}$)
\State resultTotal $\gets$ 0
\For {$i = 0$ to $n-1$}
\State start $\gets$ $i \times d$, end $\gets$ start + $d$
\State result $\gets$ 0, $\delta \gets 0$
\For {$j =$ start to end-1}
\State $\delta \gets$ $\delta$ + $X'[j] - \hat{X'}[j]$
\State result $\gets$ result + $|\delta|$
\EndFor
\State result $\gets$ result / ($d$-1)
\State resultTotal $\gets$ resultTotal + result
\EndFor
\State resultTotal $\gets$ resultTotal / $n$
\State \Return resultTotal
\end{algorithmic}
\normalsize
\end{algorithm}

\begin{algorithm}[!ht]
\small
\caption{Distance between output vectors $Y$, $\hat{Y}$}
\label{al:distance-output} 
\textbf{Inputs}: \small Strategy vectors $Y, \hat{Y}$, deck size $d$, number of bet sizes $b$ 
\begin{algorithmic}
\State resultTotal $\gets$ 0
\For {$i = 0$ to $d-1$}
\State start $\gets i \times b$; end $\gets$ start + $b$
\State result $\gets$ 0, $\delta \gets 0$
\For {$j =$ start to end-1}
\State $\delta \gets$ $\delta$ + $Y[j] - \hat{Y}[j]$
\State result $\gets$ result + $|\delta|$
\EndFor
\State result $\gets$ result / ($b$-1)
\State resultTotal $\gets$ resultTotal + result
\EndFor
\State resultTotal $\gets$ resultTotal / $d$
\State \Return resultTotal
\end{algorithmic}
\end{algorithm}

\normalsize

\section{Experiments}
\label{se:experiments}
We constructed a database of 100,000 game instances by generating random hand distributions and then computing a Nash equilibrium using the linear program formulation with Gurobi's solver~\cite{Gurobi14:Gurobi}. The na\"{\i}ve approach for constructing the distributions (of assigning uniform distributions for the players independently) is incorrect because it does not account for the fact that if one player is dealt a card then the other player cannot also be dealt that card (as is the case in real poker). We instead used a Algorithm~\ref{al:random-distribution}. We first generate the two distributions independently as in the na\"{\i}ve approach, using the procedure described in Algorithm~\ref{al:random-point}. Algorithm~\ref{al:random-distribution} then multiplies these individual probabilities together only for situations where players are dealt different cards to compute a joint distribution over the private information (these values are then normalized). The procedure could be more generally applicable beyond this setting.  We then create the cdf values from the joint distribution to be used as the inputs to the learning problem. 

\begin{algorithm}[!ht]
\caption{Generate point uniformly at random from $n$-dimensional simplex}
\label{al:random-point} 
\small
\textbf{Inputs}: dimension $n$
\begin{algorithmic}
\State $s = 0$
\For {$i = 0$ to $n-1$}
\State $a[i] \gets$ randomDouble(0,1)
\State $a[i] \gets -1 \times \log(a[i])$
\State $s \gets s + a[i]$
\EndFor
\For {$i = 0$ to $n-1$}
\State $a[i] \gets a[i] / s$
\EndFor
\State \Return a
\end{algorithmic}
\normalsize
\end{algorithm}

\begin{algorithm}[!ht]
\caption{Generate private information distribution}
\label{al:random-distribution} 
\small
\textbf{Inputs}: dimension $n$, independent distributions $x_1,x_2$
\begin{algorithmic}
\State $s = 0$
\For {$i = 0$ to $n-1$}
\For {$j = 0$ to $n-1$}
\If {i != j} 
\State next $\gets x_1[i] \times x_2[j]$
\State $x^*[i][j] \gets$ next
\State $s \gets s +$ next
\EndIf
\EndFor
\EndFor
\For {$i = 0$ to $n-1$}
\For {$j = 0$ to $n-1$}
\State $x^*[i][j] \gets x^*[i][j] / s$
\EndFor
\EndFor
\State \Return $x^*$
\end{algorithmic}
\normalsize
\end{algorithm}

\normalsize

We experimented with several data representations. The first was described above. The second uses the pdf values as the 20 features instead of the cdfs. The third separates each datapoint into 10 different points, one for each hand of player 1. Here the first 20 inputs are the cdfs as before, followed by a card number (1--10), which can be viewed as an additional 21st input, followed by the 31 strategy probabilities for that card. The fourth uses this approach with the pdf features. The 5th and 6th approaches are similar, but for the 21st input they list the cdf value of the card, not the card itself. The 7th--10th are similar to the 3rd--6th, but randomly sample a single bet size from the strategy vector to use as the output. 

\begin{figure*}[!ht]
\centering
\begin{minipage}{.49\textwidth}
\centering
\includegraphics[width=0.85\linewidth]{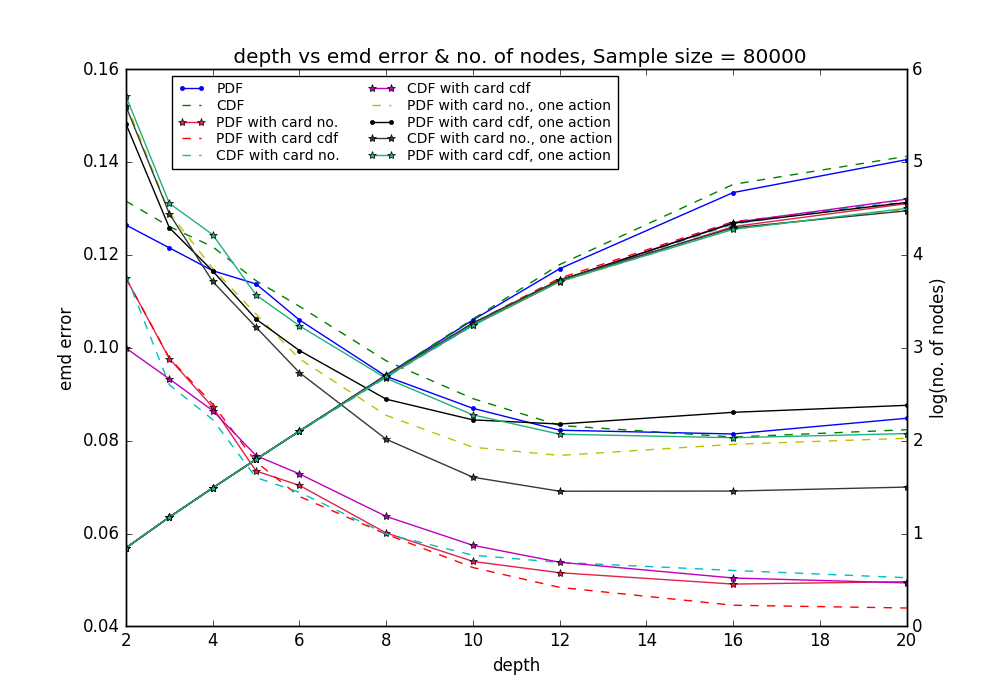}
\caption{Depth vs. number of nodes and error. The errors are the curves decreasing in depth while the number of nodes are increasing.}
\label{fi:dt-nodesanderror}
\end{minipage}
\hfill
\centering
\begin{minipage}{.49\textwidth}
\centering
\includegraphics[width=0.85\linewidth]{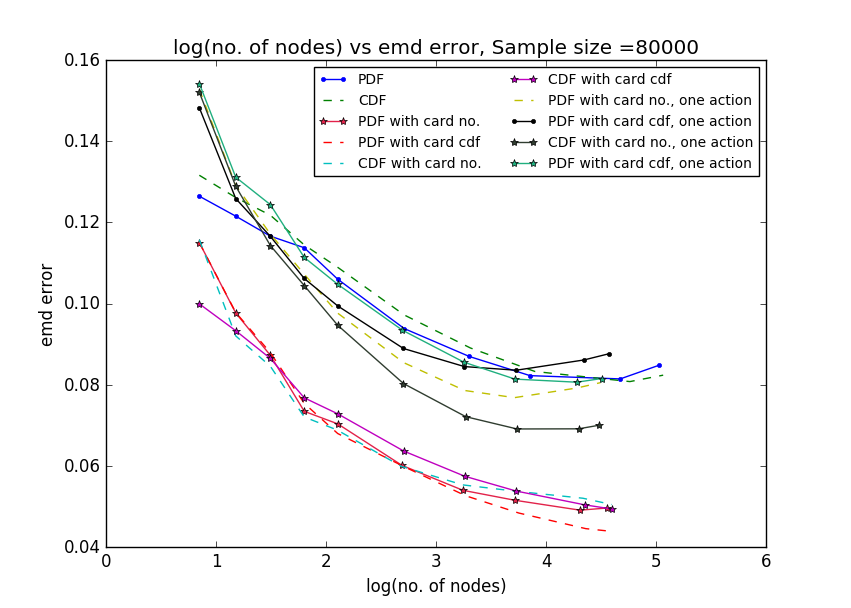}
\caption{Number of nodes vs. error in decision tree.}
\label{fi:dt-nodes}
\end{minipage}
\end{figure*}

We created decision trees using 80,000 of the games from the database, using the standard division of 80\% of the data for training and 20\% for testing (so we trained on $64,000$ games and tested on 16,000). We used Python's built in decision tree regressor function from sklearn.tree from the scikit-learn library, which we were able to integrate with our new distance metrics. 
We constructed the optimal decision tree for depth ranging from 3 up to 20. From Figure~\ref{fi:dt-nodesanderror}, we can see the errors of the optimal tree as a function of the depth, for each of the different data representations. Not surprisingly error decreases monotonically with depth; however, increasing the depth leads to an exponential increase in the number of nodes. Figure~\ref{fi:dt-nodes} shows how error decreases as a function of the (log of the) the number of nodes in the optimal decision tree.

\begin{figure*}[!ht]
\centering
\includegraphics[scale=0.24]{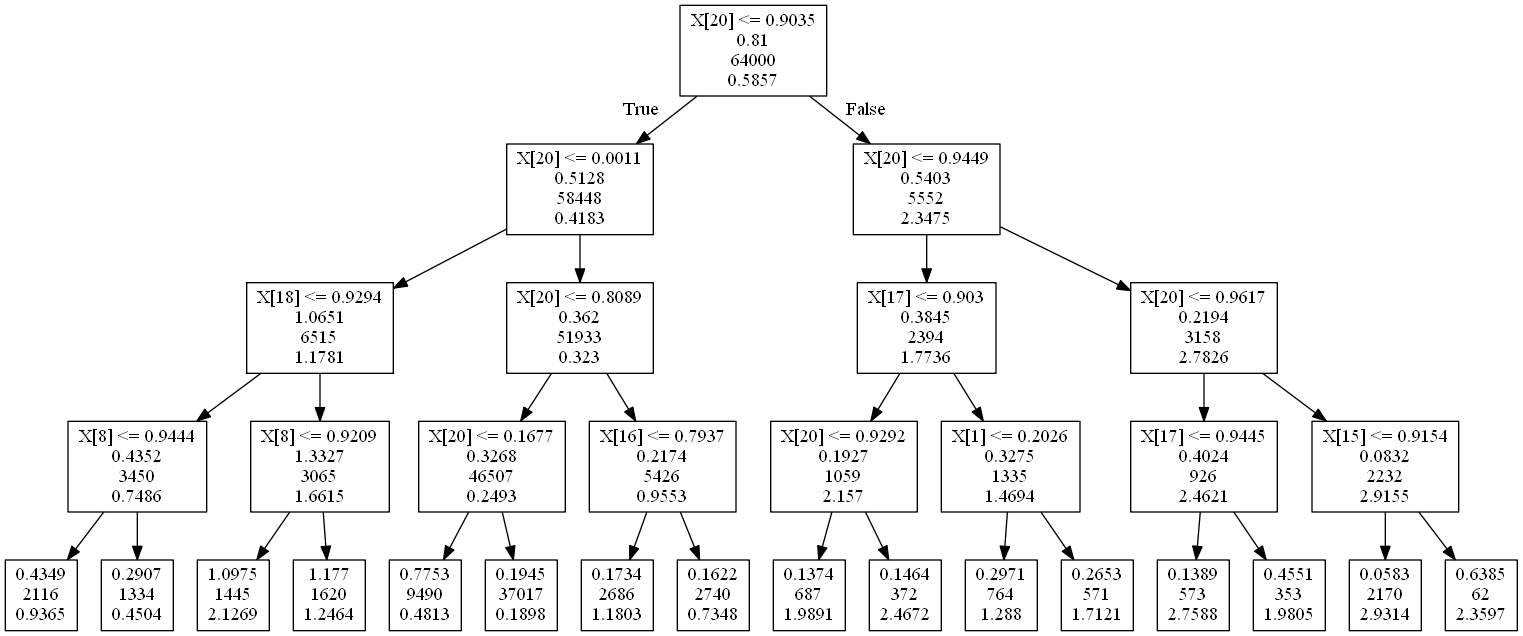}
\caption{Rules for optimal depth-4 decision tree.}
\label{fi:decision-tree}
\end{figure*}

The fourth representation, which uses the pdf values plus the cdf value of the card as the inputs, produces lowest error, and in general the approaches that output the data separately for each card produce lower errors than the ones that output full 310-length vectors. Note that if we were using full exploitability to evaluate the strategies produced then almost surely using the full 310 outputs would perform better, since they will take into account aggregate strategies across all hands to ensure the overall strategy is balanced; if we bet one size with a weak hand and a different size with a strong hand, it would be very easy for the opponent to  exploit us. In fact, for other machine learning approaches, such as $k$-nn, using the full 310 outputs performs better; but for decision trees using separate outputs for each hand leads to better branching rules which produces lower error. Note that the pdf features encode more information than the cdfs (it is possible for two games with the same cdfs to have different pdfs), so we would expect pdf features to generally outperform cdf features. 


The optimal depth-4 decision tree for the approach that uses cdf input values, cdf value for a single card, and predicts a single bet size, is shown in Figure~\ref{fi:decision-tree}. At each node the left edge corresponds to the True branch and the right to the False branch. The node entries are the variable/value branched on, the error induced due to the current branch split, the sample for the current split, and the output (i.e., bet size) predicted (leaf nodes only contain the final three). For example, the far right leaf node says that if $x[20] <= 0.9035$ is false (i.e., player 1 has a hand with cdf value $> 0.9035$), ..., if $x[15] <= 0.9154$ (i.e., the cdf value of a card 6 for player 2 is $> 0.9154$), then output a bet size of 2.3597.    

From the optimal tree, we can deduce several fundamental rules of poker strategy. The ``80-20 Rule'' is based on the branch leading to the very small bet size of 0.1898, and the ``All-in Rule'' is based on the branch leading to the large bet size of 2.9314 on the far right of the tree.

\begin{fundamental}[80-20 Rule]
If your hand beats between 20\% and 80\% of the opponent's distribution of hands, then you should always check (or make an extremely small bet).
\end{fundamental}

\begin{fundamental}[All-In Rule]
If your hand beats 95\% of the opponent's distribution of hands, and furthermore the opponent's distribution contains a weak or mediocre hand no more than 90\% of the time (i.e., it contains at least 10\% strong hands), then you should go all-in.
\end{fundamental}

Prior ``fundamental rules'' have been proposed, but often these are based on psychological factors or personal anecdotes, as opposed to rigorous analysis. For example, Phil Gordon writes, ``\textbf{Limping\footnote{In poker a ``limp'' is a play when one player matches the antes in the first play as opposed to putting in a larger bet.} is for Losers.} This is \emph{the most important fundamental} in poker---for every game, for every tournament, every stake: If you are the first player to voluntarily commit chips to the pot, open for a raise. Limping is inevitably a losing play. If you see a person at the table limping, you can be fairly sure he is a bad player.
''~\cite{Gordon11:Phil} 

\section{Conclusion}
\label{se:conclusion}

We presented a novel formulation of the problem of computing strong game-theoretic strategies that are human understandable as a machine learning problem. Traditionally computing strong strategies in games has fallen under the domain of specialized equilibrium-finding algorithms that produce massive strategy files which are unintelligible to humans. We proposed a novel formulation where the input features are the private information cdf values and the outputs are the strategy probability vectors, and we devised novel distance functions between pairs of inputs and outputs that generalize the successful earth mover's distance. We also provided a novel procedure for generating random distributions of private information, which we used to create a large database of game solutions. We created algorithms that compute strategies that can be easily implemented by humans, and deduced several new fundamental rules about poker strategy.

We note that the contributions are not specific to poker games. The model and formulation are general, and would apply to any imperfect-information game where agents are given ordered private information signals. The approaches could also apply to perfect-information games where we can generate a database of games by modifying the values of natural parameters. The approaches are also not specific to two-player zero-sum games, though they do assume that solutions can be computed for the games used in training, which can be more challenging for other game classes.

We would like to further evaluate our new distance metrics. An effective distance metric between strategies would have many potential applications. For instance, a recent algorithm for opponent exploitation computed the ``closest'' strategy to a given prior strategy that agreed with the observations, using several different distance metrics (EMD outperformed L1 and L2)~\cite{Ganzfried11:Game}.  Effective strategy distance metrics would also be useful for detecting ``bot'' agents on online sites who are playing a strategy similar to a specific known strategy. We would also like to implement full-game exploitability as a new metric to evaluate the ``cost'' of a strategy, which can be integrated with all the learning approaches.

\clearpage
\clearpage
\bibliographystyle{named}
\bibliography{F://FromBackup/Research/refs/dairefs}

\end{document}